\documentclass{aastex}
\usepackage{emulateapj5}
\usepackage{lambdabar}

\newcommand{\sci}[2]{\mbox{$#1 \times 10^{#2}$}}
\newcommand{\lambdaC}{\lambdabar_C}
\newcommand{\lnlambda}{\ln \Lambda}
\newcommand{\units}[1]{\; #1}
\newcommand{\NR}{N\!R}
\newcommand{\PFF}{P\!F\!F}
\newcommand{\vc}[1]{\mbox{\boldmath ${\bf #1}$}}    

\shorttitle{Pair Multiplicities and Pulsar Death}
\shortauthors{Hibschman and Arons}

\begin{document}
\title{Pair Multiplicities and Pulsar Death}
\author{Johann A.\ Hibschman \altaffilmark{1,2}}
\and
\author{Jonathan Arons \altaffilmark{1,2,3}}
\affil{University of California, Berkeley \altaffilmark{4}}
\altaffiltext{1}{Theoretical Astrophysics Center}
\altaffiltext{2}{Physics Department}
\altaffiltext{3}{Astronomy Department}
\altaffiltext{4}{Address correspondence to J.~A. Hibschman, Astronomy 
Dept., 601 Campbell Hall, U.C., Berkeley 94720-3411; email:
{\tt jhibschman@astron.berkeley.edu}}

\begin{abstract}
  Through a simple model of particle acceleration and pair creation
  above the polar caps of rotation-powered pulsars, we calculate the
  height of the pair-formation front (PFF) and the dominant photon
  emission mechanism for the pulsars in the Princeton catalog.  We
  find that for most low- and moderate-field pulsars, the height of
  the pair formation front and the final Lorentz factor of the primary
  beam is set by
  nonresonant inverse Compton scattering (NRICS), in the
  Klein-Nishina limit.  NRICS is capable of creating pairs over a wide
  range of pulsar parameters without invoking a magnetic field more
  complicated than a centered dipole, although we still require a
  reduced radius of curvature for most millisecond pulsars.  For
  short-period pulsars, the dominant process is curvature radiation,
  while for extremely high-field pulsars, it is resonant
  inverse Compton scattering (RICS).  The dividing point between NRICS
  dominance and curvature dominance is very temperature-dependent;
  large numbers of pulsars dominated by NRICS at a stellar temperature
  of $10^6$ K are dominated by curvature at $10^5$ K.

  Our principle result is a new determination of the theoretical
  pulsar death line.  Proper inclusion of ICS allows us to reach the
  conclusion that all known radio pulsars are consistent with pair
  creation above their polar caps, assuming steady acceleration of a
  space charge limited particle beam.  We identify a new region
  of the $P$-$\dot{P}$ diagram where slow pulsars with narrow
  radiation beams should be found in sufficiently large surveys.
  
  We also show that the injection rate of electrons and positrons into
  the Crab Nebula inferred from the polar cap pair creation model at
  the present epoch ($10^{39}$ electrons and positrons per second)
  suffices to explain the nebular X- and $\gamma$-ray emission, the
  best empirical measure of the instantaneous particle loss rate from
  any pulsar, but that the contemporary injection rate is about a
  factor of 5 below the rate averaged over the Nebula's history
  required to explain the Nebular radio emission (assuming the Nebular
  radio source is homogeneous).  It is not clear whether this
  discrepancy can be resolved by evolutionary effects and by better
  treatment of nebular inhomogeneity, or is an indication of another
  particle source in the pulsar's magnetosphere.
\end{abstract}


\maketitle

\section{Introduction}

The region above the polar caps of rotation-powered pulsars has long
been the focus of research.  From the first, e.g. \citet{goldreich69},
theoretical papers have centered on the idea that a beam of electrons
(or ions) is accelerated up from the polar cap by the electric
potential induced by the rotation of the pulsar's intense magnetic
field.  Shortly afterwards, \citet{sturrock71} suggested that this
particle beam will emit high-energy gamma-rays via curvature
radiation, which will then interact with the background magnetic field
and generate copious numbers of electron-positron pairs, forming a
pair plasma.  \citet{ruderman75} suggested that this plasma, once
formed, will act to short out the potential, preventing further
acceleration.  \citet{arons79} then showed that this scenario can
occur even when the star freely emits the particle beam.

This basic model has remained qualitatively unchanged for the past 25
years.  Recently, however, two important alterations have been made to
the theory.  First, the importance of general relativistic frame
dragging in the creation of the accelerating potential was discovered
by \citet{muslimov90, muslimov92} and later elaborated by \citet{muslimov97}. 
Second, several authors \citep{sturner95b, luo96, zhang97} have suggested
that inverse Compton scattering (ICS) is an important, perhaps the
dominant, high-energy photon emission mechanism, rather than the
curvature radiation proposed by the earlier theories.  \citet{zhang00}
used a model of a Compton-dominated pulsar including the GR
acceleration to obtain predictions for the emitted gamma-ray spectrum
but did not discuss the pair-production rate itself in any detail.

The dominant emission mechanism and accelerating potential combine to
set the height of the pair formation front (PFF), the point above the
polar cap where enough pairs have been formed to halt any further
acceleration of the beam.  Given the potential, the PFF height sets
the final Lorentz factor of the beam, determining the energy budget of
the polar-cap region.

This paper calculates, in a direct analytic manner, and for an
arbitrary pulsar, the height of the PFF in the presence of the
Muslimov and Tsygan accelerating potential and including the effects
of inverse Compton scattering.

Qualitatively, two rates combine to position the PFF: the creation
rate of photons capable of pair-producing and the rate at which those
photons can pair-produce.  Section \ref{sec:pair-production} discusses
the approximations used for the pair-production process, Section
\ref{sec:emission} discusses the emission rate of convertible photons
due to various emission mechanisms, Section \ref{sec:acceleration}
explains the model of the accelerating potential used, and Section
\ref{sec:PFF} calculates the resultant PFF height.

For simplicity, for the rest of this paper we express all energies and
temperatures in units of the electron energy, $m_e c^2 = 511
\units{keV} = \sci{5.9}{9} \units{K}$, unless otherwise specified.

\section{Pair Production}
\label{sec:pair-production}

The photons capable of pair-producing in the background dipole
magnetic field are those with energy greater than a critical value,
\begin{equation}
  \epsilon_a = \frac{32}{3} \frac{B_q}{B} \frac{\rho_e}{r_e \lnlambda}
             = 2166 \, B_{12}^{-1} P^{1/2} r_6^{-1} f_\rho 
\end{equation}
where $B$ is the local magnetic field, $B_q$ is the critical magnetic
field, $B_q = \sci{4.41}{13}\,G$, $r_e$ is the radius of the emission
point as measured from the center of the magnetic dipole, $\rho_e$ is
the radius of curvature of the magnetic field at $r_e$, and
$\lnlambda$ is treated as a constant parameter, here $\lnlambda = 20$
\citep{arons79}.  The numerical value expresses the magnetic field in
units of $10^{12}$ Gauss, $B = 10^{12} \, B_{12}\,G$, and assumes that
that the emission point is at $r_e = 10^6 r_6\,\mbox{cm}$ and that the
radius of curvature is a factor $f_\rho$ times the radius of curvature
of a dipole field on the last closed field line, $\rho_e = f_\rho
\rho_B$, $\rho_B = 4R_*/3\theta_c = \sci{9.2}{7}\, P^{1/2}\,
\mbox{cm}$, where $P$ is the pulsar spin period.  The expected
possible range of $f_\rho$ is from $f_\rho = 0.011 \, P^{-1/2}$, for
an extreme multipole field with radius of curvature equal to the
stellar radius, to $f_\rho = 1$, for a standard dipole field.

These photons (with $\epsilon > \epsilon_a$) must propagate through
the magnetic field for a distance
\begin{equation}
  \Delta s = \frac{1}{4} \frac{\epsilon_a}{\epsilon} \frac{r_e}{R_*},
  \label{eq:deltas}
\end{equation}
in units of the stellar radius $R_{*}$ before they pair-produce. 
Since the pair-production opacity is a sharply increasing function of
distance, essentially all such photons pair-produce at this distance. 
This approximation is best for low- and mid-range field strengths,
where $B < B_q/3$.  At higher fields, not only does the opacity
increase more gradually with distance, but other physical processes,
such as photon splitting \citep{harding97}, may become important. 
Most observed pulsars are in the lower regime of magnetic field.

The created pair will be in a high Landau level and will therefore
emit synchrotron radiation, bleeding off the component of its momentum
perpendicular to the local magnetic field.  The synchrotron photons
with energies above $\epsilon_a$ then pair-produce themselves,
initiating a cascade.

The total number of pairs produced by this cascade may be found by
using a modified version of the generational model due to
\citet{tademaru73}.  Each generation of pair production and subsequent
synchrotron emission lowers the maximum energy of the photon spectrum
by a factor of $\lnlambda$ \citep{arons79}.  Since the expected
synchrotron number spectrum due to one photon has a power-law index of $\nu =
-3/2$, the energy is concentrated in the high-frequency end of the
spectrum, and we may treat all the photons of the $i^{\mathrm{th}}$
generation as having energy $\epsilon_{i} = \epsilon_{i-1}/\lnlambda$.
If there is no energy loss between successive generations, this
produces a final pair multiplicity $M(\epsilon_{0}) =
\epsilon_0/\epsilon_a$ from an initial photon of energy $\epsilon_0$.

However, even in the best case, this overestimates the pair
production, since a significant fraction of the penultimate generation
of pairs will have energies less than $\epsilon_a$ and will thereby be
lost.  Given the power-law index of $\nu = -3/2$, this reduces the
number of photons available to the last generation by a factor of
$\sqrt{\lnlambda}$, giving a single-photon pair multiplicity of
\begin{equation}
  M(\epsilon_0) = 1 + \frac{1}{\sqrt{\lnlambda}}
    \frac{\epsilon_0}{\epsilon_a},
\end{equation}
namely the sum of the single pair generated by the primary photon and 
those generated by the cascade.

This pair production cascade does not take place instantaneously; it
gradually develops over the distance of $r_e/4$ it takes the lowest
energy convertible photons to pair produce.  Most pair production is
initiated by photons emitted close to the surface of the star, so
$r_e/R_{*} \approx 1$ and the length of the cascade is simply
$R_{*}/4$.  Over this span, the pair production rate is linear with
distance.  Each successive generation multiplies the number of
generated pairs, but it does so over a similarly multiplied
propagation distance.

The multiplicity of pairs produced before a height $s$ by a single
photon of energy $\epsilon$ emitted at height $s_e$ then has two
parts.  First, as long as the primary photon has enough distance to
pair produce, i.e. $s - s_e > \epsilon_a/4 \epsilon$, at least one
pair will be generated.  Second, the cascade of secondaries produced
by that pair will generate $\epsilon/(\epsilon_a \sqrt{\lnlambda})$ more
pairs, reduced by the fraction of the full $R_*/4$ available to
pair-produce:
\begin{equation}
  M(s, s_e, \epsilon) = 1 + \frac{4}{\sqrt{\lnlambda}}
    \frac{\epsilon}{\epsilon_a} (s - s_e)
  \label{eq:mult}
\end{equation}
This assumes that $s - s_e < 1/4$ and ignores the change in the
magnetic field through the cascade.  Since the magnetic field
decreases with height above the polar cap, this on-the-spot
approximation slightly over-estimates the amount of pair creation, but
it does so consistently for all of the emission mechanisms considered,
and so should not change the qualitative results.  The quality of 
these approximations will be discussed in a subsequent paper.

\section{Emission}
\label{sec:emission}

The rate at which these photons are produced depends on the emission
mechanism.  There are three primary emission mechanisms which are
important over pulsar polar caps: curvature radiation, resonant
inverse Compton scattering (RICS), and nonresonant inverse Compton
scattering (NRICS) in both the Thomson and Klein-Nishina regimes.

Each of these mechanisms produces its own characteristic radiation
spectrum.  However, since all of these spectra are roughly power-laws
harder than $\dot{N}(\epsilon) \propto \epsilon^{-1}$ until truncated
at some mechanism-dependent maximum energy, most of the emitted energy
is concentrated in photons with energy near this maximum.  We then
approximate the spectra by assuming that the entire emitted power is
converted into photons at the maximum energy, so $\dot{N}(\epsilon) =
P_X/\epsilon_{X}$, where $P_X$ is the power emitted and $\epsilon_X$
the characteristic energy of an emission mechanism $X$.

The characteristic maximum energies of these mechanisms are
\begin{eqnarray}
  \epsilon_{\NR}^{Th} & = & 2 \gamma^2 T \Delta \mu 
    \label{eq:EnrTh} \\
  \epsilon_{\NR}^{KN} & = & \gamma
    \label{eq:EnrKN} \\
  \epsilon_R & = & 2 \gamma \epsilon_B
    \label{eq:Er} \\
  \epsilon_C & = & \frac{3}{2} \frac{\lambdaC}{\rho} \gamma^3
    \label{eq:Ec}
\end{eqnarray}
where $\gamma$ is the Lorentz factor of the scattering particle,
$\epsilon_B \equiv B/B_q$, $\rho$ is the local field line radius of
curvature, $T$ is the neutron-star temperature in units of $mc^2$,
$\Delta \mu$ is the range of $\cos \theta$ from which thermal photons
arise, and $\lambdaC$ is the reduced Compton wavelength, $\lambdaC =
\sci{3.86}{-11}\,\mbox{cm}$.  Near the surface of the star, $\Delta
\mu = 1$, while at altitudes large compared to the size of the
emitting region $\Delta \mu$ decreases as an inverse-square law.  The
transition between the Thomson and the Klein-Nishina regimes of NRICS
is at $2 \gamma T \Delta \mu = 1$ or $\gamma = \sci{3.2}{3} \Delta
\mu^{-1} T_6^{-1}$. Above that $\gamma$, NRICS is in the KN limit.

The powers emitted by these mechanisms are
\begin{eqnarray}
  P_{N}^{Th} & \;\approx\; & 30.7 \, \gamma^2 T_6^4 \Delta \mu^3 \units{mc^2
    s^{-1}}, \ 2 \gamma T \Delta \mu  < 1
  \label{eq:PnrTh} \\
  P_N^{KN} &\;\approx\;& \sci{3.08}{8} \, T_6^2 \Delta \mu
    (\ln (2 \gamma T \Delta \mu) + 1)
  \units{mc^2 s^{-1}}, \nonumber \\
   & & 2 \gamma T \Delta \mu > 1 \label{eq:PnrKN} \\
  P_R &\;\approx\;& \sci{4.92}{11} \, \gamma^{-1} B^2_{12} T_6 \ln
  (\gamma T \Delta \mu \epsilon_B^{-1}) \units{mc^2 s^{-1} }, \nonumber \\
   & & \gamma T \Delta \mu \epsilon_B^{-1} > 1
  \label{eq:Pr} \\
P_C      &\;\approx\;& \sci{5.63}{-19} \, \gamma^4 \rho_8^{-2} \units{mc^2
  s^{-1}}
  \label{eq:Pc}
\end{eqnarray}
where equations (\ref{eq:PnrTh}) and (\ref{eq:PnrKN}) are from
\citet{blumen70}, equation (\ref{eq:Pr}) is from \citet{dermer90}, and
equation (\ref{eq:Pc}) is the standard form for curvature radiation,
c.f. \citet{jackson}.

Since, for our purposes, only the gross magnitudes of these power
losses are important, we treat the logarithmic terms in equations
(\ref{eq:Pr}) and (\ref{eq:PnrKN}) as constant for the rest of the
analysis.  Resonant ICS peaks at low energies, so we set $\ln (\gamma
T \Delta \mu \epsilon_B^{-1}) = 1$.  Non-resonant ICS remains
important even to high energies, so we set $\ln (2 \gamma
T \Delta \mu) = 4$, a value representative of the final Lorentz factor of the
primary beam in pulsars where NRICS sets the PFF.

For particles moving at moderate Lorentz factors ($\gamma <
\sci{4.6}{6}\, T_6^{1/2} P^{1/4}$), resonant and nonresonant
inverse Compton scattering are the dominant energy-loss mechanisms
\citep{sturner95}, while curvature dominates at higher Lorentz
factors.  For the rest of this section, we adopt a model in which the
entire surface of the star is hot, giving $\Delta \mu \approx 1$ for
the active low-altitude region.  Only NRICS is strongly dependent on
$\Delta \mu$, and the effects of a hot polar cap on that process are
discussed in Section \ref{sec:nrics}.

At moderate $\gamma$,
NRICS generally overwhelms RICS.  The Klein-Nishina NRICS process
clearly produces photons with the largest possible energy, namely the
entire energy of the scattering particle.  If RICS is to be
significant, thermal photons must be scattered into resonance with
the field, so $\gamma T > \epsilon_B$, therefore we also have
$\epsilon_N^{Th} > \epsilon_R$.  In both Klein-Nishina and Thomson
regimes, then, the NRICS photons have a higher energy than the RICS
photons, and will thus pair produce after propagating a shorter
distance.  If the emission rates of the two processes were equal, the
higher-energy NRICS photons would set the location of the PFF, simply
due to their faster conversion into pairs.

At low $\gamma$, the RICS emission rate is greater than NRICS,
but the RICS rate fades quickly at higher $\gamma$.  RICS photons
first pair produce when $\epsilon_R > \epsilon_a$, or at a minimum
Lorentz factor of
\begin{equation}
  \gamma_R^{min} = \frac{1}{2} \frac{\epsilon_a}{\epsilon_B} \approx
    \sci{4.8}{4} \, B_{12}^{-2} P^{1/2}.
\end{equation}
If NRICS produces more photons even at this minimum Lorentz factor, it
will clearly dominate RICS at all later $\gamma$s.  At
$\gamma_R^{min}$, NRICS operates in the Thomson regime if $2
\gamma^{min}_R T < 1$, or $B_{12} > 4.02 T_6^{1/2} P^{1/4}$, and in
the Klein-Nishina regime otherwise.  Most pulsars, then, are in the
Klein-Nishina regime by the time RICS can generate pairs.

Dividing the emitting powers by the respective characteristic energies
gives an estimate for the photon production rates for Klein-Nishina
NRICS and RICS,
\begin{eqnarray}
\dot{N}_N^{KN} & \approx & \sci{1.54}{9}  \, \gamma^{-1} T_6^2 \Delta \mu  \, s^{-1} \\
\dot{N}_R      & \approx & \sci{1.08}{13} \, \gamma^{-2} B_{12} T_6 \, s^{-1}
\end{eqnarray}

Given these rates, NRICS in the Klein-Nishina regime will produce more
convertible photons than RICS ($\dot{N}_{N}^{KN} > \dot{N}_R$)
whenever $\gamma > \sci{7.0}{3}\,B_{12} T_6^{-1} \Delta \mu^{-1}$.
Clearly, the RICS emission is strongest at lower $\gamma$.  However,
even if we evaluate the emission rates at $\gamma^{min}_R$, NRICS
produces more photons whenever
\begin{equation}
  B_{12}^{KN} < 1.9 \, P^{1/6} T_6^{1/3} \Delta \mu^{1/3}
\end{equation}

For pulsars with magnetic fields less than $B_{12}^{KN}$, RICS has no
chance of determining the pair formation front; at all beam $\gamma$'s
high enough for RICS to generate pairs, NRICS has a higher emission
rate.

For fields greater than $B_{12}^{KN}$, the dominant process is still
unclear.  RICS will only dominate the creation of pair-producing
photons for a small range of $\gamma$, and those photons will still
take longer to pair-produce than the NRICS photons.  As the primary
beam accelerates through this range, RICS will only dominate for a
short distance.  If RICS produces enough pairs in that distance, then
it will set the PFF, but this is likely true only for high-field
pulsars.  This expectation is borne out by the more detailed
calculations in Section \ref{sec:PFF}.

Curvature emission, on the other hand, produces photons with energy
large enough to pair produce only when $\gamma$ is on the order of
$10^7$ or higher.  At this beam energy, both NRICS and RICS emission
rates are negligible; the question becomes whether the beam has a
chance to accelerate to such high Lorentz factors in the presence of
radiative losses.  In Section \ref{sec:PFF}, we find that curvature
radiation is only important for the short-period pulsars, due to their
higher voltage.

\section{Acceleration Model}
\label{sec:acceleration}

In this paper, we adopt a simple model of the accelerating potential,
based on the general relativistic frame dragging first proposed by
\citet{muslimov92}, and similar to low-altitude limits of the
potential derived by \citet{muslimov97}.  In this model, the charge
difference arises because, close to the star, frame dragging reduces
the effective rotation frequency $\Omega_{eff}$ to less than the
observed value.  The density of the beam extracted from the star
equals the Goldreich-Julian charge density at the surface, but is
insufficient to cancel the charge at higher altitudes as
frame-dragging weakens and the effective rotation frequency approaches
the asymptotic flat-space value.  To maintain simplicity, we neglect
any additional effects of GR, such as those of \citet{gonthier94};
these effects introduce no more than 10\% corrections to our results.

The effective rotation rate is $\Omega_{eff} = (1 - \kappa_g x^{-3})
\Omega$, where $x \equiv r/R_*$ is the dimensionless distance from the
center of the star and $\kappa_g \equiv 2 G I/c^2 R_*^3 \approx
0.15\,I_{45} R_6^{-3}$, with $I$ the moment of inertia of the star.
The effective Goldreich-Julian density is then $\eta_{GJ}(x) =
\eta_{GJ}^0 (1 - \kappa_g x^{-3})$, where $\eta_{GJ}^0$ is the charge
density with frame dragging neglected, $\eta_{GJ}^0 = -\vc{\Omega}
\cdot \vc{B}/2 \pi c$.

For a more detailed form of the accelerating potential, see
\citet{muslimov97}.  For our purposes, we divide the potential into two
portions: the low-altitude quadratic potential at the surface of the
star, representing the fringe fields, and a high-altitude linear
potential.  The linear form of the high-altitude potential assumes
that the distance above the star, $s \equiv x - 1$, is less than 1; $s
< 1/2$ suffices.  The low-altitude fringe fields occupy $s < \theta_c
\ll 1$.  We assume the cross section of the polar flux tube to be
round with cross sectional area $\pi \theta_c^2 R_*^2 x^3$, as is
consistent with observations of pulsar beaming morphology
\citep{lyne88, kramer98}.

Due to the conducting boundary of the surface of the polar cone, the
high-altitude form of the field is simply that of a slightly flared
cylindrical charged waveguide operated below cutoff.  Assuming that
the beam density matches the Goldreich-Julian density at the stellar
surface, the difference in charge density at high altitudes is $\Delta
\eta(s) = -3 \kappa_g \eta_{*}^0 (B(x)/B_*) s$, where $\eta_*^0 \equiv
\eta_{GJ}^0(R_*)$.  Since this is above the fringe field zone, the
induced electric field is almost cylindrically radial, giving a
potential of $\Phi = \pi \theta_c^2 R_*^2 (1 - w^2) (B_* \Delta
\eta(s)/B(x)) $, where $\theta_c$ is the colatitude of the last closed
field line, $\theta_c = \sqrt{\Omega R_*/c}$, $w$ is the fractional
position across the polar cap, $w \equiv \theta_*/\theta_c$, and
$\theta_*$ is the colatitude of the footprint on the stellar surface
of the current field line.

Using the potential drop across the polar cap, $\Phi_{cap} =
\theta_c^2 \Phi_0$, where $\Phi_0 \equiv 4 \pi \eta_*^0 R_*^2$ is a
pole-to-equator estimate of the potential, the potential of a field
line at high altitudes becomes
\begin{equation}
  \Phi_{high}(s) = \frac{3}{4} \kappa_g \Phi_{cap} s
    (1 - w^2).
\end{equation}

Numerically, $\Phi_0 = \sci{1.26}{17}\,B_{12} P^{-1}\,V$ and
$\Phi_{cap} = 2.63\times10^{13} \, B_{12} P^{-2} \, V$ where we have
neglected the $\cos \alpha$ inclination term as it is of order 1. 
This produces an order of magnitude stronger acceleration for $P \sim
1\,\mbox{sec}$ than the previous models of \citet{arons79}, which
depended on the curvature of the field lines to generate the
starvation electric field.  For $P \lesssim 10$ ms, however, the frame
dragging and field curvature acceleration mechanisms are comparable.

The low-altitude field can be treated as that of a plane-parallel charged
slab, so we expect the potential to increase quadratically.  The
low-level charge density is slightly different from the
Goldreich-Julian density.  Following the method of \citet{arons79}, we
choose this density difference and the height of the transition
between the low and high so that the potential and electric field are
continuous.

This gives a low-altitude potential of
\begin{equation}
  \Phi_{low} = \frac{1}{2} \Phi_0 \kappa_g s^2
    (\sqrt{6} \theta_c (1 - w^2)^{1/2} - s)
\end{equation}
where the transition between the low and high forms of the potential
occurs at $s = \sqrt{3/2}\,\theta_c (1 - w^2)^{1/2}$.

To reduce this to an even more convenient form, we drop the cubic
term from the low-altitude potential and simply extend the quadratic
component until it meets the high-altitude linear potential.  This occurs
at
\begin{equation}
  s_1 \equiv \frac{1}{2} \sqrt{\frac{3}{2}} \theta_c (1 - w^2)^{1/2}
   = \sci{8.87}{-3} P^{-1/2} (1 - w^2)^{1/2}.
  \label{eq:s1}
\end{equation}
If we express the particle Lorentz factor in terms of the scaled
height, $t \equiv s / s_1$, we obtain
\begin{equation}
  \begin{array}{rcll}
    \gamma_{low}(t)  & = & \gamma_1 t^2, & t < 1 \\
    \gamma_{high}(t) & = & \gamma_1 t,   & t > 1
  \end{array}
\end{equation}
where
\begin{equation}
  \gamma_1 \equiv \frac{3}{8} \sqrt{\frac{3}{2}} \kappa_g \theta_c^3
    \frac{e \Phi_0}{mc^2} = \sci{5.14}{4}\,B_{12} P^{-5/2}
    (1 - w^2)^{3/2}.
  \label{eq:gamma1}
\end{equation}
Although we have dropped the smooth transition in electric field from
low to high as an approximation, we are still using that underlying
model to set the coefficient.  The approximation is valid as long as
the acceleration is not radiation reaction limited for particle
energies $\approx \gamma_1$, so that $\gamma \propto \Phi$, as is true
of all of our models.

This acceleration is strong enough that we can ignore the effects of
radiative losses on the beam particles due to ICS, so this
discontinuous electric field causes no difficulties.  Since the power
produced by ICS declines with increasing Lorentz factor,
once the beam has accelerated a short distance above the stellar
surface, the ICS losses become negligible.  Curvature radiation becomes
very efficient at high Lorentz factor, so if the beam reaches a
Lorentz factor of order a few times $10^7$, curvature radiation will
hold it there.  We will return to this issue when we discuss curvature
radiation in more depth.

For the remainder of the paper, we simply drop the dependence on $w$
from the potential, in effect examining the activity on a typical
field line, where neither $w$ nor $1-w$ is small.

\section{Pair Formation Front}
\label{sec:PFF}

The primary beam from the star will accelerate due to the starvation
electric field until the beam has produced a pair plasma dense enough
to short out the field.  Several authors have located this pair
formation front (PFF) at the point where photons emitted by the
primary beam first reach an optical depth of one, with regards to the
pair-creation opacity.  This heuristic was first suggested by
\citet{ruderman75} for curvature-dominated cascades, and in that case,
it is valid.  For curvature radiation, further acceleration of
particles only increases both the energy and number of emitted
photons, resulting in a sharp front of pair production.

However, if inverse Compton radiation is the primary source of
convertible photons, this approximation breaks down.  As the primary
beam accelerates, the energy of the ICS photons increases, but their
number drops.  Rather than the copious cascades produced by curvature
radiation, where the PFF is a sharp transition, ICS tends to produce
more gradual cascades where a small number of high-energy particles
set off a lengthy cascade of pair production.  For inverse Compton
processes, then, it is not guaranteed that the potential will be
shorted out as soon as the first pair is produced.

To properly calculate the location of the pair formation front in the
face of relatively sparse photon production, we have to track not only
the height at which photons of a certain energy are emitted and the
distance those photons travel before pair producing, but also the
number of such photons produced and the number of secondary pairs
produced by the resulting synchrotron cascade.  In short, we have to
integrate the pair production rate until we reach some threshold.
Since inertial frame dragging induces a fractional difference of the
beam density from the Goldreich-Julian density of approximately
$\kappa_g$, we set the location of the pair formation front by
equating the number of pairs created per primary particle to
$\kappa_g$.

\subsection{PFF Position}

As discussed in Section \ref{sec:emission}, for each emission
mechanism we approximate the emitted number spectrum by assuming that
all of the power goes into creating photons at the respective
characteristic maximum energy:
\begin{equation}
  \dot{N}_X(\epsilon) \approx \frac{P_X(\gamma)}{\epsilon_X(\gamma)}
    \delta(\epsilon - \epsilon_X(\gamma))
\end{equation}
where $X \in \{C, \NR, R\}$, and $P_X(\gamma)$ and
$\epsilon_X(\gamma)$ are again the total power and the maximum energy
emitted by particles with Lorentz factor $\gamma$ due to curvature
(C), nonresonant ICS (NR), or resonant ICS (R) emission.

Using the cascade multiplicity (\ref{eq:mult}), the total number of
pairs produced by one primary particle before a height $s$ may be
divided into a portion due to the primary photons themselves, $N_{\pm
  1}$, and a portion due to the subsequent cascade of secondaries,
$N_{\pm 2}$, where
\begin{eqnarray}
  N_\pm(s) & = & N_{\pm 1}(s) + N_{\pm 2}(s) \label{eq:s_integral} \\
  N_{\pm 1}(s) & = & \int_{s_{min}}^{s_{max}} ds' R_* \frac{P_X(s')}{c
    \epsilon_X(s')} \label{eq:s_integral1} \\
  N_{\pm 2}(s) & = & \frac{4}{\sqrt{\lnlambda}} \int_{s_{min}}^{s_{max}}
    ds' R_* \frac{P_X(s')}{c \epsilon_a} (s-s')
    \label{eq:s_integral2}
\end{eqnarray}
where $s_{min}$ and $s_{max}$ are the first and last heights at which
the primary beam emits photons capable of pair producing at or below
$s$.  We neglect the variation of $\epsilon_{a}$ with $s'$, making an
on-the-spot approximation.

These heights are the values of $s_e$ which satisfy
\begin{equation}
  \Delta s(\epsilon_X(s_e)) = s - s_e
\end{equation}
with $\Delta s$ from equation (\ref{eq:deltas}).  For fast
acceleration, $s_{min} \ll s$ and $s - s_{max} \ll s$, so these
limits are approximately given by
\begin{equation}
 \epsilon_X(s_{min}) \approx \frac{\epsilon_a}{4 s}
\end{equation}
\begin{equation}
  s_{max} \approx s - \frac{1}{4} \frac{\epsilon_a}{\epsilon_X(s)}
\end{equation}
The lower height $s_{min}$ is the distance required to accelerate
particles to energies capable of pair producing in a distance $s$,
while $s_{max}$ is $s$ reduced by the distance it would take a photon
emitted at $s$ to pair produce.  By the time a pulsar beam has reached
an altitude where the PFF forms, it has typically reached a high
energy, so $s_{max} \approx s$.  The form of $s_{min}$, however, will
depend on the regime of the potential in which it falls.  Since the
potential switches from its low-altitude form to its high-altitude
form at $s=s_1$, it is easiest to calculate in terms of the scaled
distance, $t \equiv s/s_1$.

\subsubsection{Non-resonant ICS}
\label{sec:nrics}

%
%

As the beam pulled from the stellar surface accelerates to high
$\gamma$, it will pass from the Thomson regime of NRICS into the
Klein-Nishina regime.  Since this acceleration is extremely strong,
this transition will occur close to the stellar surface, and we can
simply use the Klein-Nishina form of NRICS for these calculations. 
For NRICS in the KN regime, $P_{\NR}(\gamma)$ is only logarithmically
dependent on $\gamma$, so we treat it as effectively constant.  Since
the power is then independent of the Lorentz factor, the NRICS cascade
is insensitive to the details of the accelerating potential.

Furthermore, since the NRICS process generates comparatively few
photons, it can only achieve high pair multiplicities through extended
cascades of secondaries.  Using the NRICS power, equation
(\ref{eq:PnrKN}), in the secondary portion of the multiplicity,
equation (\ref{eq:s_integral}), produces a total multiplicity as a
function of height of
\begin{equation}
    N_{\pm \NR}(t) = 2 \frac{P_{\NR}}{\epsilon_{a} \sqrt{\lnlambda}} 
      \frac{s_1^2 R_{*}}{c} (t - t_{min})^{2}
\end{equation}
The critical energy is $\epsilon_{\NR} = \gamma$, so if we define
$t'_{min} \equiv \epsilon_a/4 s_1 \gamma_1 t$, then $t_{min} =
t_{min}^{\prime \, 1/2}$ if $t'_{min} < 1$ and $t_{min} = t'_{min}$
otherwise.

Typically, however, the PFF only forms at $t \gg t_{min}$, so we can
neglect $t_{min}$ above.  The distance to achieve a pair multiplicity
of $\kappa$ is then
\begin{equation}
    t = \left(\frac{\kappa \epsilon_a \sqrt{\lnlambda}}{2 P_{\NR}} 
    \frac{c}{s_1^2 R_{*}}\right)^{1/2}
\end{equation}
The PFF takes place where $\kappa = \kappa_g$, or
\begin{equation}
  t_{\PFF, \NR} = 13.4 \, B_{12}^{-1/2} P^{3/4} T_6^{-1} f_\rho^{1/2}
    \label{eq:t_pff_nr}
\end{equation}
assuming, as before, $\kappa_g = 0.15$.  In terms of the stellar
radius, this is
\begin{equation}
  s_{\PFF, \NR} = 0.119 \, B_{12}^{-1/2} P^{1/4} T_6^{-1} f_\rho^{1/2}
    \label{eq:s_pff_nr}
\end{equation}

For most pulsars, this is well above the regime of the fringe fields,
$t \approx 1$.  Likewise, it is still within the linear region of the
potential ($s < 1$), provided that $B > \sci{2}{10}$ G.  Very young 
and energetic pulsars may have $t_{\PFF} < 1$, as discussed in 
Section \ref{sec:discussion}.  The above formalism remains valid 
regardless.

If only the polar cap is hot, the flux of thermal X-rays is roughly
halved by $t = 1$, reducing the number of pairs produced.  If we model
a hot polar cap by applying the same technique as above, but
attenuating the ICS source at $t > 1$ by $1/t^2$, we find that the
pair formation front occurs at
\begin{equation}
  t_{\PFF, \NR}^{cap} = 45.0 \, B_{12}^{-1} P^{3/2} T_6^{-2} f_\rho
    \label{eq:t_pff_nr_cap}
\end{equation}
or
\begin{equation}
  s_{\PFF, \NR}^{cap} = 0.399 \, B_{12}^{-1} P T_6^{-2} f_\rho
    \label{eq:s_pff_nr_cap}
\end{equation}
As expected, the PFF occurs at higher altitudes if only the polar cap
is hot.

\subsubsection{Resonant ICS}

In general, resonant inverse Compton scattering depends on the large
resonant cross section to compete with nonresonant scattering.  The
RICS photons are typically lower energy than the NRICS photons and
produce less of a cascade.  As such, the pair production of the
primaries dominates that of the secondaries, and we use equation
(\ref{eq:s_integral1}) to compute the multiplicity.

The steps of computing the PFF height are the same as for NRICS, using
the primary-only integral (\ref{eq:s_integral1}) rather than equation
(\ref{eq:s_integral2}) and $t'_{min} = \epsilon_a/8 \epsilon_B s_1
\gamma_1 t$.

Since the emission rate depends on particle $\gamma$, the two forms of
the accelerating potential yield different results.  There are three
major regimes: where the pair formation front forms in the
low-altitude region, with $t, t_{min} < 1$, where the PFF forms in the
upper region due to photons emitted in the lower, with $t_{min} < 1$
but $t > 1$, and where the entire process takes place in the
high-altitude potential, with $t, t_{min} > 1$.

In practice, RICS is only important when it can create pairs in the
low-altitude potential with $t_{min} < 1$.  If that is so, the
number of pairs produced as a function of distance in both cases is,
to leading order in $t_{min}$,
\begin{equation}
  N_{\pm R}(t) = \frac{1}{3} \frac{P_R(\gamma_1)}{2 \gamma_1 \epsilon_B} 
   \frac{s_1 R_*}{c} t_{min}^{-3}
\end{equation}
which gives a PFF height of
\begin{equation}
  t_{\PFF,R} = 1350 \, B_{12}^{-7/3} P^{1/2} T_6^{-2/3} f_\rho
    .
    \label{eq:t_pff_r}
\end{equation}
or
\begin{equation}
  s_{\PFF,R} = 12.0 \, B_{12}^{-7/3} T_6^{-2/3} f_\rho
    .
  \label{eq:s_pff_r}
\end{equation}

\subsubsection{Curvature Radiation}

Curvature radiation differs from both ICS processes in that, as the
beam Lorentz factor increases, both the emission rate and the average
photon energy increase.  This increase creates a sharp transition in
the pair production rate once the particle $\gamma$ reaches the
minimum required to pair-produce.  Due to this quick generation of a
copious pair plasma, both the $N_{\pm 1}$ (\ref{eq:s_integral1}) and
$N_{\pm 2}$ (\ref{eq:s_integral2}) forms of the multiplicity lead to
nigh-identical results for the PFF height.  In practice, we find that
the approximation derived from $N_{\pm 2}$ better fits the numerical
solution, so we will concentrate on that case.

The curvature power (\ref{eq:Pc}) and critical energy (\ref{eq:Ec})
give for $t,\,t_{min}\,<\,1$,
\begin{equation}
  N_{\pm C}^{low} = \frac{2}{45}
    \frac{P_C(\gamma_1)}{\epsilon_a \sqrt{\lnlambda}} \frac{s_1^2 R_*}{c}
    (t^{10} - 10 t t_{min}^9 + 9 t_{min}^{10})
    \label{eq:n_c_low}
\end{equation}
and for $t, t_{min} > 1$,
\begin{equation}
  N_{\pm C}^{high} = \frac{2}{15} 
    \frac{P_C(\gamma_1)}{\epsilon_a \sqrt{\lnlambda}} \frac{s_1^2 R_*}{c}
    (t^6 - 6 t t_{min}^5 + 5 t_{min}^6)
    \label{eq:n_c_high}
\end{equation}
The mixed case is unimportant, due to the strong $t$-dependence of the
multiplicity.

As before, $t_{min}$ is the first point at which the beam has reached
a sufficient Lorentz factor for the curvature photons to pair-produce
by $t$,
\begin{equation}
  t_{min} \approx
   \left(\frac{\epsilon_a \rho}{6 \lambdaC s_1 t}\right)^{1/3} =
   894 \, t^{-1/3} B_{12}^{-4/3} P^3 f_\rho^{2/3} 
\end{equation}

Since the power produced by curvature radiation increases strongly
with increasing $\gamma$, the pair multiplicity is dominated by the
emission from the upper end of the range, so we can obtain a simple
approximation by assuming $t_{min} = 0$.  As most pulsars are in the
$t_{\PFF} > 1$ regime, except for the shortest-period millisecond
pulsars, we concentrate on high-$t$, giving
\begin{equation}
  t_{\PFF,C}^{high} \approx 98.2 B_{12}^{-5/6} P^{25/12} f_\rho^{1/2}
    \quad \mbox{if}\quad t_{\PFF,C}^{low} > 1
    \label{eq:t_pff_c}
\end{equation}
When compared to the result of numerically solving for $t$, this is
only correct to within a factor of 2--3.  This discrepancy arises
because once curvature emission is active, it very rapidly shorts out
the field, so $t_{\PFF}$ is close to $t_{min}$.  Neglecting $t_{min}$,
while giving a straightforward expression for the PFF height,
introduces an appreciable error.  The results given in subsequent
sections use the full numerical solution to the polynomials
(\ref{eq:n_c_low}) and (\ref{eq:n_c_high}) for $t_{\PFF,C}$; equation
(\ref{eq:t_pff_c}) should be considered a rough estimate.

In terms of stellar radius, the high-altitude limit becomes
\begin{equation}
  s_{\PFF,C}^{high} \approx 0.678 \, B_{12}^{-5/6} P^{19/12} f_\rho^{1/2}
    .
    \label{eq:s_pff_c}
\end{equation}
This depends primarily on the polar cap potential, as $s_{\PFF} \propto
\Phi_{cap}^{-5/6} P^{-1/12}$.  Physically, obtaining any pair
production from curvature emission requires high potentials and beam
Lorentz factors.

These results are inaccurate, however, for the millisecond pulsars,
because radiation reaction prevents the primary beam from accelerating
to the high Lorentz factors needed to generate photons capable of
pair-producing in the low magnetic fields.  The high-altitude
accelerating electric field is
\begin{equation}
  E = \frac{\gamma_1}{s_1} \units{\frac{mc^2}{e R_*}}.
\end{equation}
The acceleration due to this field is balanced by the curvature power
loss at a Lorentz factor of
\begin{equation}
  \gamma_{bal} = \sci{2.26}{7} \, B_{12}^{1/4} P^{-1/4} f_\rho^{1/2}.
\end{equation}
Setting the typical curvature photon energy at $\gamma_{bal}$ to
$\epsilon_a$, $\epsilon_C(\gamma_{bal}) = \epsilon_a$, and solving for
the magnetic field yields
\begin{equation}
  B_{12} = 0.500 \, P f_\rho^{2/7}
  \label{eq:rad_reaction}
\end{equation}
At fields lower than this, radiation reaction will prevent curvature
radiation from creating pairs.  We cannot yet conclude that at higher
fields, curvature photons will create pairs, merely that radiation
reaction will not prevent it.

If the magnetic field has a radius of curvature of $\rho = R_*$, or
$f_\rho = 0.011 \, P^{-1/2}$, then the limiting magnetic field becomes
\begin{equation}
  B_{12} = 0.137 \, P^{6/7},
\end{equation}
which would allow more pulsars to pair-produce via curvature
radiation.

\subsection{Beam Lorentz Factor}

The PFF is located at
\begin{equation}
  t_{\PFF} = \min(t_{\PFF,\NR},\, t_{\PFF,R},\, t_{\PFF,C}).
\end{equation}

Since the beam stops accelerating at this point, its final Lorentz
factor is
\begin{eqnarray}
  \gamma_{beam} = \gamma_1 t_{\PFF}^2, & \mbox{if} & t_{\PFF} < 1
    \label{eq:g_pff_low} \\
  \gamma_{beam} = \gamma_1 t_{\PFF},   & \mbox{if} & t_{\PFF} > 1
    \label{eq:g_pff_high}
\end{eqnarray}

Analytically, the estimated final Lorentz factors for each emission
process, assuming $t_{\PFF} > 1$, are
\begin{eqnarray}
  \gamma_{\NR} & = & \sci{6.90}{6} \, B_{12}^{1/2} P^{-7/4} T_6^{-1}
    f_\rho^{1/2}
    \label{eq:gamma_nr}
    \\
  \gamma_{R}  & = & \sci{6.95}{7} \, B_{12}^{-4/3} P^{-2} T_6^{-2/3}
    f_\rho
    \label{eq:gamma_r}
    \\
  \gamma_{C} & = & \sci{5.05}{6} \, B_{12}^{1/6} P^{-5/12}
    f_\rho^{1/2}.
    \label{eq:gamma_c}
\end{eqnarray}
Since the smallest $t_{\PFF}$ sets the gap height, the dominant
mechanism is the one with the lowest predicted beam $\gamma$.

\subsection{Polar Cap Heating}
\label{sec:polar_heating}

Basic electrodynamics requires that a fraction of the positrons
created above the PFF be reversed and accelerated back down onto the
surface of the star, heating it.  In the curvature-starvation
acceleration model of \cite{arons79}, this flux is $\theta_c^{2}
n_{GJ} c$, while in the frame-dragging-starvation model, we expect a
reversed flux on the order of $\kappa_{g} \theta_{c} n_{GJ} c$, where
$n_{GJ} \equiv \eta_{GJ}/e$ is the fiducial Goldreich-Julian number
density, see \citet{harding98, zhang00, zhang00b}.

Since these reversed particles are accelerated through the same
potential as the primary beam, they strike the polar cap with a
Lorentz factor $\gamma_{\PFF}$.  The downward energy flux is then
\begin{equation}
  \phi = \gamma_{\PFF} mc^{2} f_{rev} n_{GJ} c
\end{equation}
where $f_{rev}$ is the fraction of the Goldreich-Julian flux that is
reversed.  The energy is then radiated away by a surface black body,
yielding a polar cap temperature
\begin{eqnarray}
  T_{cap} & = & \left(\frac{\gamma_{\PFF} mc^{2} f_{rev} n_{GJ} c}{ 
    \sigma_{SB}}\right)^{1/4}
    \label{eq:temp_cap} \\
          & = & \sci{3.9}{6} \, B_{12}^{1/2} P^{-3/4}
                \left(f_{rev}\frac{\gamma_{\PFF} mc^{2}}{e \Phi_{cap}}
                \right)^{1/4} \, K.
\end{eqnarray}

Since the value of $\gamma_{\PFF}$ for the ICS processes itself
depends on the stellar temperature, we can solve equation
(\ref{eq:temp_cap}) for the self-consistent cap temperature and, from
that, derive the PFF height due to each process.  Setting $f_{rev} =
\kappa_g \theta_c$, this yields cap temperatures of
\begin{eqnarray}
  T_{6,\NR} & = & 0.730 \, B_{12}^{1/6} P^{-5/12} \, f_{\rho}^{1/6} 
    \label{eq:cap_temp_nr} \\
  T_{6,R} & = & 1.38 \, B_{12}^{-1/14} P^{-3/4} \, f_{\rho}^{3/14}
    \label{eq:cap_temp_r} \\
  T_{6,C} & = & 0.758 \, B_{12}^{7/24} P^{-23/48} \, f_{\rho}^{1/8}
\end{eqnarray}
where we have assumed that $t_{\PFF} > 1$, using equation
(\ref{eq:t_pff_r}) for RICS, the attenuated NRICS expression, equation
(\ref{eq:t_pff_nr_cap}), and equation (\ref{eq:g_pff_high}) for the
resultant Lorentz factor.  If $t_{\PFF} < 1$, the appropriate
corresponding equations would be (\ref{eq:t_pff_r}),
(\ref{eq:t_pff_nr}), and (\ref{eq:g_pff_low}).  For curvature, only
the high-altitude form is important, equation (\ref{eq:t_pff_c}).  The
low-altitude expressions yield
\begin{eqnarray}
  T_{6,\NR}^{low} & = & 0.919 \, B_{12}^{1/6} P^{-5/12} \, f_{\rho}^{1/6}
    \label{eq:cap_temp_nr_low} \\
  T_{6,R}^{low} & = & 5.13 \, B_{12}^{-1/2} P^{-9/16} \, f_{\rho}^{3/8}.
    \label{eq:cap_temp_r_low}
\end{eqnarray}
The low-altitude forms are most appropriate for the youngest, highest
potential pulsars.

These temperatures correspond to PFF heights of
\begin{eqnarray}
  s_{\PFF,\NR,pc} & = & 0.749 \, B_{12}^{-4/3} P^{11/6}
    \, f_{\rho}^{2/3}
    \label{eq:s_pff_nr_heat} \\
  s_{\PFF,R,pc} & = & 9.66 \, B_{12}^{-16/7} P^{1/2} \, f_{\rho}^{6/7}
    \label{eq:s_pff_r_heat} \\
  s_{\PFF,\NR,pc}^{low} & = & 0.129 \, B_{12}^{-2/3} P^{2/3} \, f_{\rho}^{1/3}
    \label{eq:s_pff_nr_heat_low} \\
  s_{\PFF,R,pc}^{low} & = & 4.92 \, B_{12}^{-2} P^{3/8} \, f_{\rho}^{3/4}
    \label{eq:s_pff_r_heat_low}
\end{eqnarray}
Since the curvature PFF height does not depend on the temperature, it
remains as given in equation (\ref{eq:s_pff_c}).


The regimes in which these expressions apply are somewhat complex.
The predicted PFF height depends on whether the pulsar is in the $t <
1$ or $t > 1$ regime.  For each of RICS and NRICS, if $t_{\PFF}^{low}
\equiv s_{\PFF}^{low}/s_1 < 1$, then the low-altitude variant is
appropriate, otherwise the high-altitude.  The shortest PFF height
then determines the dominant mechanism and the temperature of the
polar cap.

However, if the stellar temperature, as derived from a cooling model,
is higher than this predicted temperature, then the polar heating is
irrelevant and the normal whole-star PFF heights apply instead,
equations (\ref{eq:s_pff_nr}) and (\ref{eq:s_pff_r}).


\subsection{Final Pair Multiplicity}

Although the beam acceleration stops at $t_{\PFF}$, the primary
particles will continue to produce pairs even after the PFF.  After
the PFF, cascades fall into two general categories: opacity-bounded
and energy-bounded.  Opacity-bounded cascades are primarily limited by
the decline of the magnetic field with distance from the star; to
first order, the beam Lorentz factor remains constant.  Energy-bounded
cascades, by comparison, are limited by the amount of energy in the
beam capable of being converted into pairs.  Here the beam loses
energy faster than the magnetic field drops.

Most pulsars dominated by inverse Compton processes are opacity
bounded.  By the time the beam has reached the PFF, it has typically
accelerated to a Lorentz factor large enough that further Compton
losses are negligible.  Pulsars dominated by curvature emission,
however, are energy-bounded.  Once the beam has a Lorentz factor high
enough to make pairs via curvature radiation, radiative losses quickly
slow the beam once the accelerating potential has been shorted out.

For NRICS from a hot star, the final number of pairs generated may
then be approximated by counting the number of pairs produced by a
constant-$\gamma$ beam,
\begin{equation}
  \kappa_{\NR} = \frac{1}{\sqrt{\lnlambda}} \int_0^\infty ds
    \frac{R_*}{c}\frac{P_{NR}}{\epsilon_a(s)}.
\end{equation}
Since $\epsilon_a(s) = \epsilon_a^0 (1 + s)^{-7/2}$, this becomes
\begin{equation}
  \kappa_{\NR} = \frac{2}{5} \frac{P_{NR}}{\epsilon_a^0 \sqrt{\lnlambda}}
    \frac{R_*}{c} = 2.1 \, B_{12} P^{-1/2} T_6^2 f_\rho^{-1}.
  \label{eq:kappa_nr_whole}
\end{equation}

If only the polar cap is hot, this is reduced by dilution to
\begin{equation}
  \kappa_{\NR}^{cap} = 2 \frac{P_{NR}}{\epsilon_a^0 \sqrt{\lnlambda}}
    \frac{s_1 R_*}{c} = 0.094 \, B_{12} P^{-1} T_6^2 f_\rho^{-1},
  \label{eq:kappa_nr}
\end{equation}
using the same approximate attenuation model ($\propto t^{-2}$ for 
$t>1$).

For RICS, the total number of pairs generated is dominated by photons
which are emitted below the PFF, but which are absorbed above it.  The
minimum height for pair production is $t_R^{min} = (\epsilon_a / 2
\gamma_1 \epsilon_B)^\alpha$, where, as before, $\alpha = 1$ for
$t_R^{min} > 1$ and $\alpha = 1/2$ otherwise.

The number of pairs produced by the primary photons is comparable to
that produced by the cascade, so the multiplicity is
\begin{equation}
  \kappa_R = \int_{t_R^{min}}^{t_R^{min}} dt \frac{s_1 R_*}{c}
    \frac{P_R(\gamma(t))}{2 \gamma(t) \epsilon_B} \left(1 +
    \frac{1}{\sqrt{\lnlambda}} \frac{2 \gamma(t) \epsilon_B}{\epsilon_a}
    \right).
\end{equation}
If $t_R^{min} < 1$, the total pair production is dominated by the
region below $t = 1$, so $t_R^{max} = 1$; otherwise, the pair
production is truncated by the decline of the magnetic field on the
scale of $R_*$ and $t_R^{max} \approx s_1^{-1}$.

This gives two expressions for the final multiplicity, the low-altitude
estimate, appropriate when $B_{12} \gtrsim P$,
\begin{equation}
  \kappa_R^{low} = \sci{7.5}{-4} \, B_{12}^{7/2} T_6 f_\rho^{-3/2}
  \label{eq:kappa_r_low}
\end{equation}
and the high-altitude estimate, appropriate when $B_{12} \lesssim P$,
\begin{equation}
  \kappa_R^{high} =
    0.0013 \, B_{12}^2 P^{3/2} T_6 f_\rho^{-1}
      \left(1 + 
      \frac{4.8 + 3 \ln B_{12} - 2.5 \ln P}{\sqrt{\lnlambda}}
      \right)
  \label{eq:kappa_r_high}
\end{equation}

If curvature sets the PFF, the cascade is energy-bounded, and the
total number of pairs is roughly the overshoot energy of the beam,
converted into pairs with energy $\epsilon_a$.  If $\gamma^{min}_C$ is
the minimum Lorentz factor required to pair produce, the total number
of pairs expected is
\begin{equation}
  \kappa_C = \frac{\gamma_C^{\PFF} - \gamma^{min}_C}{\epsilon_a
    \sqrt{\lnlambda}}
\end{equation}
where
\begin{equation}
  \gamma_C^{min} = \left(\frac{2}{3} \frac{\rho}{\lambdaC} \epsilon_a
    \right)^{1/3} = \sci{1.51}{7} \, B_{12}^{-1/3} P^{1/3} f_\rho^{2/3}.
\end{equation}
This expression for $\kappa_C$ depends critically on the slight
difference between the beam Lorentz factor and $\gamma_C^{min}$.  As
such, this form cannot be further simplified while maintaining
accuracy.

\subsection{Death Lines}
\label{sec:death_line}

Pulsar death occurs when pair-production is too weak to generate the
multiplicity of $\kappa_g$ required to short out the accelerating
potential.  This sets a minimum potential for each process,
namely
\begin{eqnarray}
  \Phi_{NR}^{death} & = &  \sci{6.0}{13} \, P^{-5/8} f_\rho^{1/2} \units{V}
\\
  \Phi_{R,low}^{death}  & = & \sci{1.1}{14} \, P^{-57/32} f_\rho^{3/8}
    \units{V}
\\
  \Phi_{R,high}^{death} & = & \sci{2.6}{14} \, P^{-43/18}
    f_\rho^{11/27} \units{V}
\end{eqnarray}
where we have used the multiplicity equations (\ref{eq:kappa_nr}),
(\ref{eq:kappa_r_low}), and (\ref{eq:kappa_r_high}), along with the
corresponding heated cap temperatures, equations
(\ref{eq:cap_temp_nr}) and (\ref{eq:cap_temp_r}).  If the star itself
is hotter than the heated caps, the stellar temperature predicted by
the cooling model used should be substituted into the multiplicity
equations, rather than the self-heating model.

The corresponding death line for curvature radiation is best computed
by simply finding the minimum $\Phi$ such that the primary beam produces
photons capable of pair-producing by $s = 1$, after which point the
acceleration saturates.  Including the decline of the magnetic
field and the increase in the radius of curvature with increasing
distance from the star gives a minimum potential of
\begin{equation}
  \Phi_{C}^{min} = \sci{8.95}{13} \, P^{-1/4} f_\rho^{1/2} \units{V}
\end{equation}
Radiation reaction, however, also sets a lower limit on the required
field.  Converting the magnetic field limit from equation
(\ref{eq:rad_reaction}) into potential units gives
\begin{equation}
  \Phi_{C}^{death} = \max (\sci{8.95}{13} \, P^{-1/4} f_\rho^{1/2},\;
    \sci{1.32}{13} \, P^{-1} f_\rho^{2/7} ) \units{V}.
\end{equation}

\begin{figure*}
  \plotone{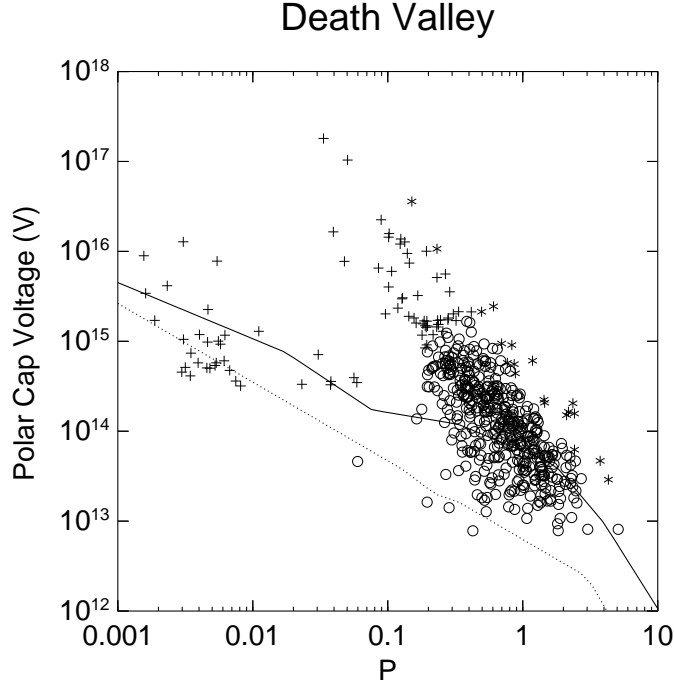}
  \caption{Pulsar death lines for a self-heated polar cap.  The solid
     line is the death line for a dipole magnetic field, while the
     dotted line assumes a $\rho = R_*$ multipole field.  Pulsars, in
     the dipole case, dominated by NRICS are marked with circles,
     those dominated by RICS are marked with stars, and those
     dominated by curvature are marked with crosses.}
  \label{fig:death_lines}
\end{figure*}

The resulting death lines are shown in terms of the polar cap
potential in Figure \ref{fig:death_lines}, for both the normal dipole
field and a $\rho = R_*$ configuration.  As low-to-mid $P$, NRICS sets
the death lines, although the dip near $P=0.1$ is due to the effects
of curvature radiation.  At high $P$, RICS sets the death line instead.

Many of the pulsars near $P=1$ are between the two death lines.  This
indicates that either non-dipolar components are important in these
pulsars, or that the temperature of these objects is larger than what
the self-heating model predicts.  We consider non-dipolar effects,
namely an offset dipole, to be the most likely explanation.

\subsection{Particle Injection into the Crab Nebula}

Study of the emission from the plerions surrounding pulsars can
provide empirical constraints on the pair multiplicity $\kappa $. The
best studied case is the Crab Nebula, energized by its central pulsar,
which has a rotation period of 33 msec and an inferred magnetic dipole
moment $\mu = 3.6 \times 10^{30} $ cgs. For this pulsar, our results
yield $\kappa \approx 5 \times 10^4 $. The Goldreich-Julian rate of
elementary charge loss is $\dot{N}_{GJ} = 2 \Omega^2 \mu /ec
= 1.8 \times 10^{34} \; {\mathrm s}^{-1}$.  The pulsar's geometry is
such that the primary polar cap current extracted by the starvation
electric field is made of electrons. Then the total rate of injection
of electrons and positrons into the nebula implied by our model is
$\dot{N}_{e^+ + e^-} = 2\kappa \dot{N}_{GJ} \approx 1.8 \times
10^{39} {\mathrm s}^{-1}$.

The nebular X-ray emission from the equatorial torus \citep{hester95}
constrains the electron plus positron injection rate from the pulsar
into the visible equatorial region to be approximately $(1-2) \times
10^{37}$ particles ${\mathrm s}^{-1}$
\citep[J.~Arons \& A.~Spitkovsky, in preparation]{kennel84, gallant94}.
The flow feeding the equatorial region fills a
latitudinal sector with full angular width of approximately $2 \delta
\sim 20^\circ $, assuming the flow from the wind termination shock to
the visible torus is purely radial. If the pair outflow from the
pulsar is spherically symmetric, our model of the multiplicity implies
$\sim 3 \times 10^{38}$ electrons and positrons per second are being
fed into the equatorial X-ray emitting region, well more than enough
to explain the X-ray observations.

In reality, both non radial divergence of the post shock flow in the
nebula and the likely decline of the pair flux toward the boundaries
of the polar flow (the region of the polar cap outflow which feeds the
equatorial) both suggest that our estimate of the particle flux {\it
  into the X-ray torus} is an upper limit. Therfore we conclude that
polar cap pair creation in the Crab pulsar is adequate to fully
explain the present day particle injection rate measure by X-ray
observations of the Crab Nebula. Other plerions will be considered
elsewhere.

By contrast, it is not so clear that polar cap pair creation can
explain the {\it total} particle injection rate into the Crab. It has
long been known that the radio emission from the Nebula requires a
particle injection rate, {\it averaged over the Nebula's history}, of
approximately $10^{40} {\mathrm s}^{-1}$ ({\it e.g.} \citet{rees74}),
an estimate probably uncertain to a factor of 3.  Our calculations
yield a total injection rate {\it today} about a factor of 5 below the
average rate required to explain the radio emission, assuming no
Nebular evolution.

Whether this discrepancy is a serious problem is unclear. The estimate
of the average rate of pair injection assumes the Nebula to be a
homogeneous synchrotron source - clearly, it is not. The pair creation
rate probably was higher in the past, when the pulsar was rotating
somewhat more rapidly, so long as the star was cool enough for pair
creation to operate on the curvature dominated branch shown in Figure
\ref{fig:tsuruta_model}, as is the case today. On the other hand,
adiabatic losses as the nebula expands might increase the required
average rate. An additional magnetospheric source of particles, such
as an outer gap, may exist. These more sophisticated nebular modeling
and global magnetospheric questions are outside the scope of our
investigation, and we leave this subject with the remark that polar
cap pair creation {\it may} be adequate to explain the total particle
injection rate into this well known synchrotron source, but the case
is not as clear as it is for understanding the injection of the
particles that create the high energy synchrotron emission.  The
application of our models to other pulsar-plerion systems is also a
topic for a separate investigation.

\section{Discussion}
\label{sec:discussion}

A broad survey, such as this, best answers broad questions about the
population.  With these results, we seek to answer two questions,
namely ``Which gamma-ray emission mechanism dominates in which region
of pulsar space?''  and ``How well does this model explain the observed
distribution of pulsars in $P$-$\dot{P}$ space?''.

\begin{figure*}
  \plotone{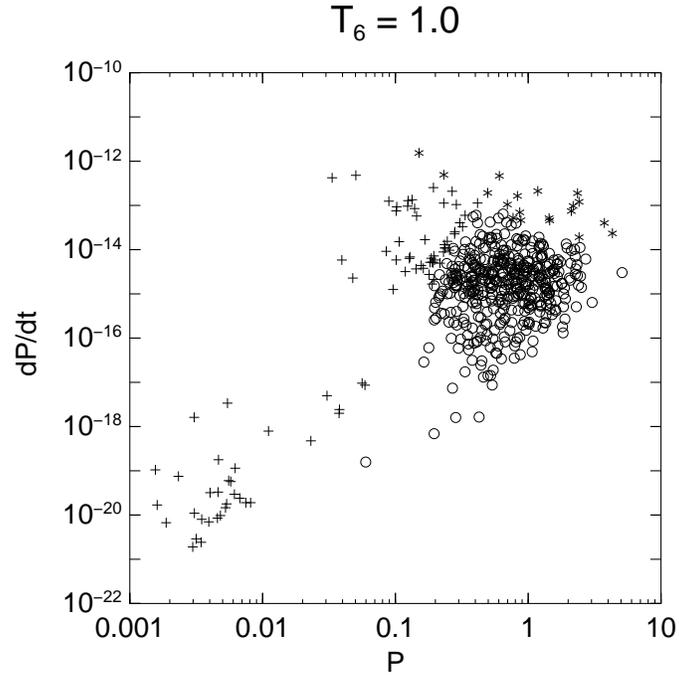}
  \caption{$P-\dot{P}$ diagram showing which pulsars have the PFF
  height set by curvature (crosses), NRICS (circles), and RICS
  (stars)}
  \label{fig:p_p_dot}
\end{figure*}

\begin{figure*}
  \plotone{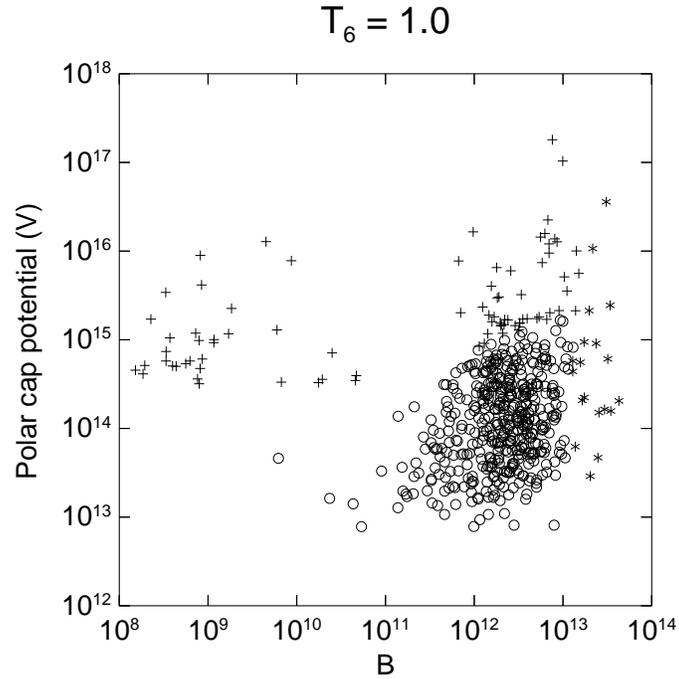}
  \caption{$B-\Phi_{cap}$ diagram showing which pulsars have the PFF
   height set by curvature (crosses), NRICS (circles), and RICS
   (stars).  Curvature dominates at high $\Phi_{cap}$, while RICS
   dominates high $B$.}
  \label{fig:b_vs_phi}
\end{figure*}

For any pulsar, the smallest of $t_{\PFF,R}$, $t_{\PFF,\NR}$, and
$t_{\PFF,C}$ sets the height of the pair formation front.  For a
whole-star temperature of $10^6$ K, assuming no special field
configuration, the pulsars of the Princeton pulsar catalog \citep{taylor93} are
categorized as shown in Figure \ref{fig:p_p_dot}.  Figure
\ref{fig:b_vs_phi} shows this same dataset in terms of the more
physical variables of surface magnetic field and polar cap voltage.
This categorization depends on the assumed temperature and size of the
thermally emitting region, but several basic conclusions can still be
drawn.  The largest-field pulsars tend to be dominated by RICS, the
largest-voltage by curvature radiation, and the rest by NRICS.

As can be seen in Figure \ref{fig:b_vs_phi}, the important boundaries
are between curvature and NRICS at high voltage and between NRICS and
RICS at high magnetic field.  Comparing the predicted PFF heights due
to RICS and NRICS yields
\begin{equation}
  \frac{t_{\PFF, \NR}}{t_{\PFF, R}} = 0.0099 \, B_{12}^{11/6} P^{1/4}
    T_6^{-1/3} f_\rho^{-1/2}
    .
\end{equation}
The boundary between NRICS and RICS is almost completely set by the
magnetic field.  RICS will dominate only for pulsars with very high
fields, $B_{12} > 12.4 \,P^{-3/22} T_6^{2/11}$.

Similarly, comparing the PFF heights due to curvature and NRICS yields
\begin{equation}
  \frac{t_{\PFF,C}^{high}}{t_{\PFF,\NR}} = 7.3 B_{12}^{-1/3} P^{4/3} T_6
    ,
\end{equation}
Whenever this ratio is less than one, the PFF is set by curvature
emission; in this regime, RICS is never important.  Thus, curvature
dominates whenever $P < 0.22 \,B_{12}^{1/4} T_6^{-3/4}$.  This
boundary is less sensitive to the magnetic field than that between
RICS and NRICS, but depends more strongly on the temperature.

Perhaps the more interesting question is where this theory can
comfortably generate enough pairs to short out the accelerating
potential, assuming that pair creation is required for pulsar
emission.  Since many pulsars are dominated by NRICS, and NRICS depends
strongly on the size and temperature of the thermally emitting region,
both the cooling model chosen for the pulsar and the degree of polar-cap 
heating are extremely important.

\begin{figure*}
  \plotone{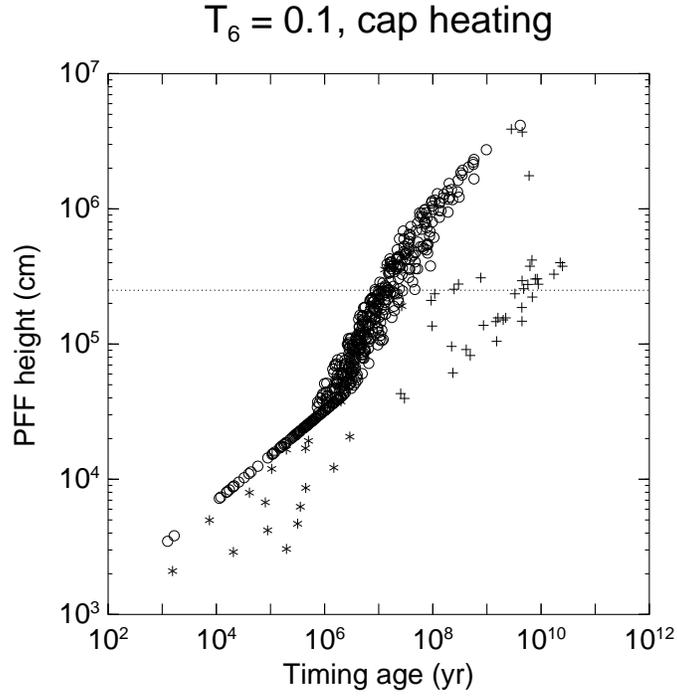}
  \caption{PFF height vs. timing age, using a self-consistent hot 
    polar cap.  Pulsars dominated by NRICS are marked with circles,
    those dominated by RICS are marked with stars, and those dominated by
    curvature are marked with crosses.  Above the dotted line at
    $s_{\PFF} = 0.25 \, R_*$, polar-cap NRICS fails to create sufficient pairs
    to halt acceleration.}
  \label{fig:polar_heating}
\end{figure*}

The case of a cool ($10^5$ K) star with the self-consistent polar cap
temperatures derived in Section \ref{sec:polar_heating} is shown in
Figure \ref{fig:polar_heating}.  Using the polar cap NRICS
multiplicity, equation (\ref{eq:kappa_nr}), we find that the cascade
cannot produce sufficient pairs to short out the electric field if the
predicted PFF is above $s_{\PFF} = 0.25$.  Given this criterion, this
model succeeds for pulsars with timing ages less than about $10^7$
years, but fails for older pulsars.

\begin{figure*}
  \plotone{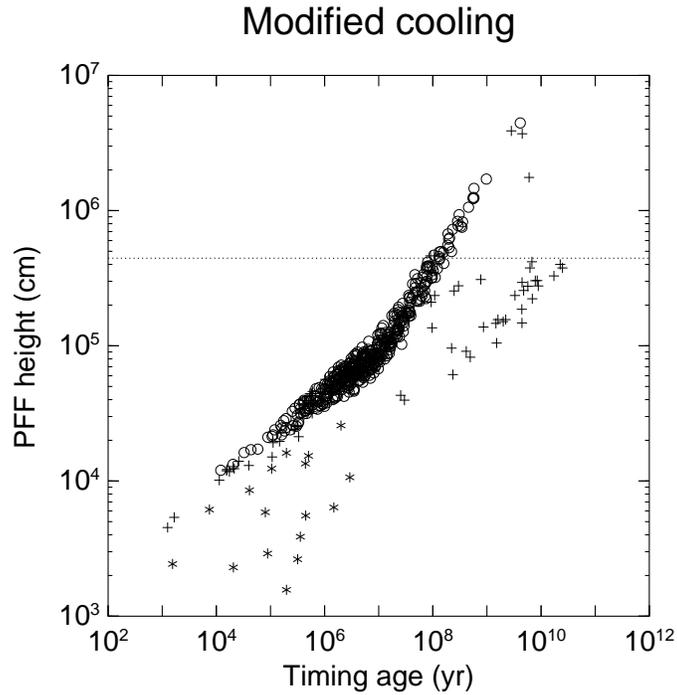}
  \caption{PFF height vs. timing age, using a temperature model 
    derived from the magnetic cooling results of \citet{tsuruta98}.
    Pulsars dominated by NRICS are marked with circles,
    those dominated by RICS are marked with stars, and those dominated by
    curvature are marked with crosses.  Above the dotted line at
    $s_{\PFF} = 0.45 \, R_*$, NRICS fails to create sufficient pairs
    to halt acceleration.}
  \label{fig:tsuruta_model}
\end{figure*}

Figure \ref{fig:tsuruta_model} shows the results of adopting a
cooling model derived from \citet{tsuruta98},
which includes the effects of the magnetic field on thermal
conductivity, but no polar cap heating.  Since heat conduction is
easier along magnetic field lines than perpendicular to them, the
polar regions remain hotter than the standard cooling of an
unmagnetized pulsar, out to ages greater than $10^6$ years.  Full 2D
calculations have only been extended to ages of several million years,
but the polar regions show no sign of rapid cooling up to that point.
For our model, we assume that more rapid cooling, with a power-law
slope of $-0.375$ begins at $10^7$ years.  This power law represents
the effects of internal heating; with no additional heating, the field
would decay exponentially beyond this point.

The precise temperature model used is
\begin{equation}
  \begin{array}{ll}
    \log T = 6.65  - 0.1 \log \tau,  & \tau < 10^7 \units{yr} \\
    \log T = 8.575 - 0.375 \log \tau,& \tau > 10^7 \units{yr}
  \end{array}
\end{equation}
where $\tau$ is the spindown age of the star.  This is a generous
model, in that it predicts a hot polar cap to late ages, but it
demonstrates the strong effect of the cooling model on the
pair-creation process.

Using the NRICS multiplicity, equation (\ref{eq:kappa_nr_whole}), we
find that the cascade cannot produce sufficient pairs to short out the
electric field if the predicted PFF is above $s_{\PFF} = 0.45$.  The
combination of magnetic cooling and internal heating allows pulsars to
remain active out to timing ages of $10^8$ years, without further
modifications.

The notable exceptions in both of these cases are the oldest pulsars
and the millisecond pulsars.  The millisecond pulsars, for the most
part, require particles to reach a Lorentz factor so large that their
acceleration is truncated by radiation-reaction before reaching the
threshold for pair-production, while the old pulsars simply do not
provide enough voltage and field strength to pair-produce.  These
pulsars could all be explained by invoking a magnetic field more
complicated than a simple star-centered dipole, however.  Due to the
success of the rotating vector model for pulsar polarization
\citep{radhakrishnan69}, we believe an offset dipole to be more
plausible than simply adding an arbitrary mixture of higher-order
multipoles.

\begin{figure*}
  \plotone{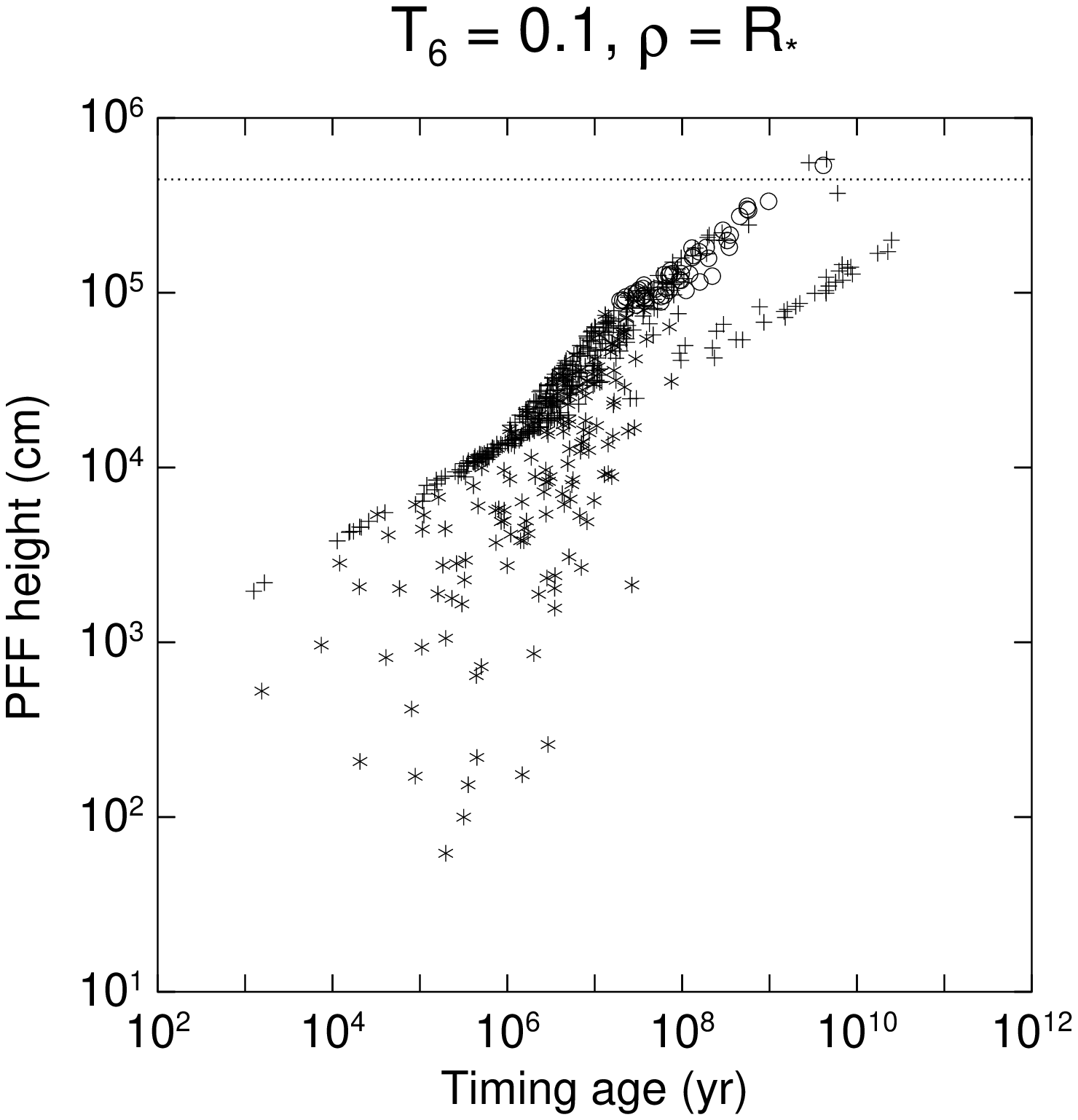}
  \caption{PFF height vs. timing age, assuming a modified magnetic 
    field such that the field line radius of curvature is $10^6$ cm.
    Pulsars dominated by NRICS are marked with circles,
    those dominated by RICS are marked with stars, and those dominated by
    curvature are marked with crosses.  Above the dotted line at
    $s_{\PFF} = 0.45 \, R_*$, NRICS fails to create sufficient pairs
    to halt acceleration.}
  \label{fig:rho_star}
\end{figure*}

An offset dipole could both increase the surface field strength by
moving the center of the dipole closer to the surface and reduce the
effective radius of curvature through gravitational bending of photon
orbits.  Decreasing the radius of curvature to $R_*$ allows almost all
pulsars to generate sufficient pairs, as shown in Figure
\ref{fig:rho_star}.  Any increase in the effective surface field would
produces similar results, when combined with the effects of
gravitational bending of the photon orbits (J.~Arons, in preparation).

\begin{figure*}
  \plotone{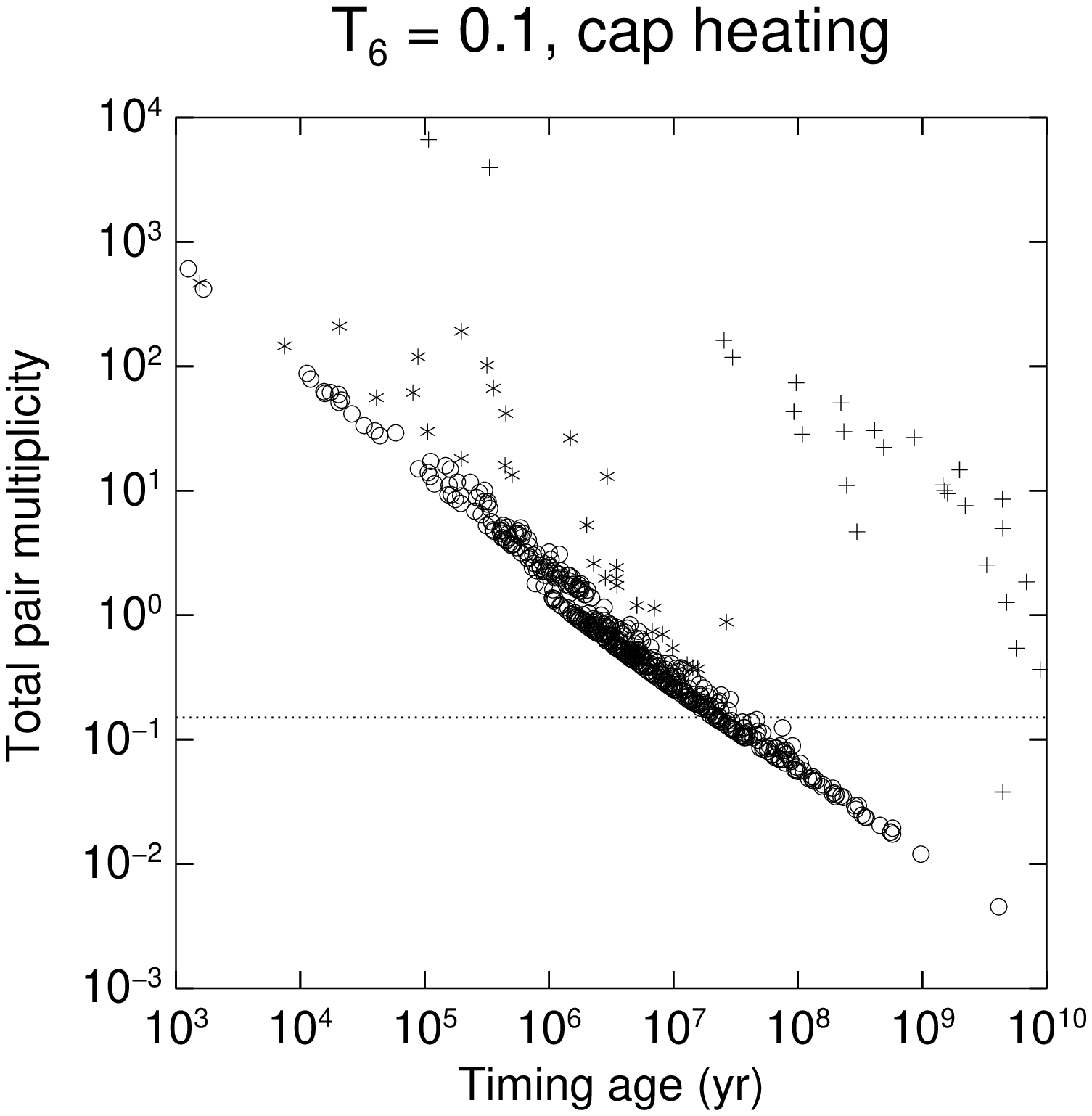}
  \caption{Final pair multiplicity, $\kappa$, using a self-consistent hot 
    polar cap.
    Pulsars dominated by NRICS are marked with circles,
    those dominated by RICS are marked with stars, and those dominated by
    curvature are marked with crosses.}
  \label{fig:mult}
\end{figure*}

For a conservative model of a self-heated polar cap, the predicted
final pair multiplicity is shown in Figure \ref{fig:mult}.  If ICS
photons create the PFF, the typical multiplicity is low, in the range
of 1--100 for the bulk of pulsars, much lower than the $10^3$
commonly assumed.  If curvature photons stop the acceleration,
however, such high multiplicities can be reached.

Since the calculation of the expected multiplicity includes the decay
of the magnetic field with distance, the multiplicity gives a good
estimate of which pulsars can form the PFF, as discussed in Section
\ref{sec:death_line}.  The total multiplicity shows a steady decline
with timing age, dropping to 0.1 near $10^7$ years, below which the
pair creation is too sparse to form the PFF.

\begin{figure*}
  \plotone{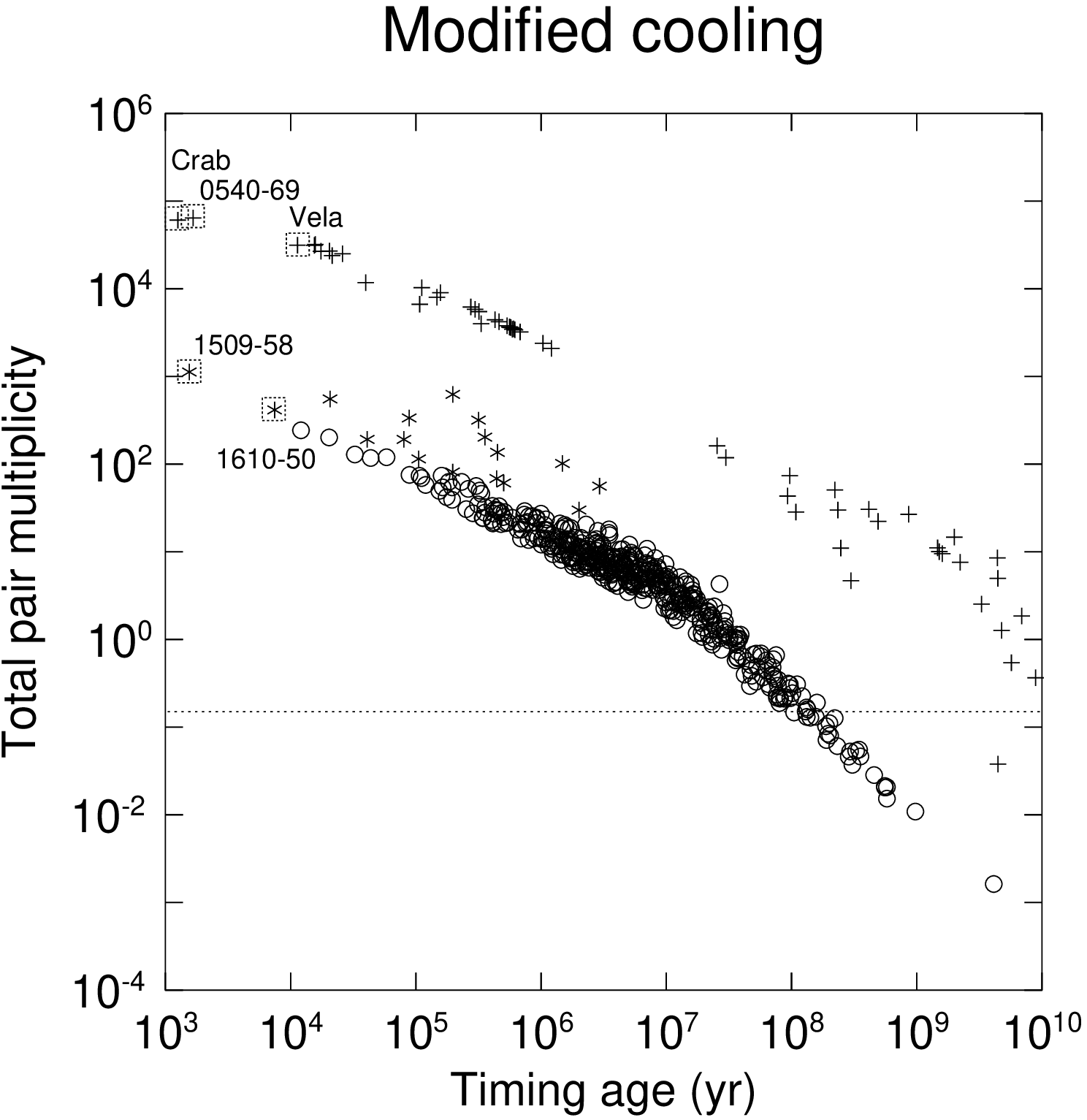}
  \caption{Final pair multiplicity, $\kappa$, using a temperature model 
    derived from the magnetic cooling results of \citet{tsuruta98}.
    Pulsars dominated by NRICS are marked with circles,
    those dominated by RICS are marked with stars, and those dominated by
    curvature are marked with crosses.  The youngest five pulsars are
    explicitly labelled.}
  \label{fig:mult_tsuruta}
\end{figure*}

In Figure \ref{fig:mult_tsuruta}, we show the pair multiplicity using
the Tsuruta temperature model and no cap heating.  The increased stellar
temperature keeps the pair multiplicity expected due to NRICS between
10--100 out to a few tens of million years, while the lack of polar
cap heating results in curvature dominance and correspondingly high
multiplicities for some of the youngest pulsars.  If polar cap heating
were included, NRICS would truncate beam acceleration for most of
those pulsars, dropping the multiplicity back to the $100-1000$ range.

\section{Conclusions}

Even without considering the competition between the various emission
processes, the most robust result of this work is that most pulsars
can generate enough pairs to short out the starvation electric field
without invoking a complex dipole structure, if the polar cap
temperature is over $10^6$ K.  However, without additional heating,
the surface temperature drops below this value after c.\ $10^6$
years, making older pulsars problematic.  For the hotter stars,
nonresonant inverse Compton scattering easily creates pairs in the
simple dipole case, unlike the earlier models of pure-curvature
cascades
\citep{arons79}.  Like any cooling neutron
star, pulsars should produce observable thermal X-rays, with some
enhancement due to the small, but hot, polar cap.

Due to the effects of ICS, the expected pair multiplicities are lower
than those expected from curvature cascades, in the range of 1--100,
in contrast to the multiplicities of $10^3$ or higher produced by
curvature cascades.  Despite the increase of the ICS cross-sections at
low energy, the primary beam of particles extracted from the pulsar
still accelerates to Lorentz factors of $10^4$ or higher, due to the
lag between the emission of pair-producing photons and their actual
conversion into pairs.

The improved death line calculations reveal a regime of
pair-production available at high magnetic fields and large periods,
first predicted by \cite{arons98}.  At large periods, the polar cap is
small and the accelerating electric field weak.  The small cap
prevents NRICS from producing enough pairs, if only the cap is hot,
but the slow acceleration allows RICS to do so easily.  The small
polar cap also produces a narrow beam, making these objects difficult
to detect.  The recent discovery of PSR J2144-3933 by \cite{young99} appears to
confirm these results, as does the more recent theoretical treatment
by \cite{zhang00b}, which should encourage further searches in this
region of parameter space.

Since each of the different emission processes dominates in different
regimes of pulsar space, the radio emission process must be
insensitive to the precise way in which the pair plasma was formed.  The
comparatively low pair-production multiplicities found constrain the
radio emission process by limiting the available particle densities.

More detailed studies are, of course, required to understand the
expected pair and gamma-ray spectra from these stars.  For any given
pulsar, more precise calculations are required to compute the beam
$\gamma$ and the PFF height as functions of distance from the magnetic
axis, but the general trends illuminated here should hold.

Our research on pulsars was supported in part by NSF grant AST 9528271
and NASA grant NAG 5-3073, and in part by the generosity of
California's taxpayers.

\end{document}